\documentclass[%
 reprint,
superscriptaddress,
 amsmath,amssymb,
 aps,
]{revtex4-2}
\usepackage{physics}
\usepackage{bbold}
\usepackage{amsmath}
\usepackage{latexsym}
\usepackage{graphicx}
\usepackage{amsthm}
\usepackage{hyperref}
\usepackage{lipsum}
\usepackage[english]{babel}
\usepackage{slashed}
\usepackage[compat=1.1.0]{tikz-feynman}
\usepackage{appendix}
\usepackage{subfig}
\usepackage{multirow}
\usepackage{mathtools}
\usepackage{subfig}
\usepackage{pgfplots}
\pgfplotsset{compat=1.15}
\usepackage{mathrsfs}
\usetikzlibrary{arrows}
\definecolor{ffffff}{rgb}{1,1,1}
\definecolor{qqqqff}{rgb}{0,0,1}
\definecolor{qqwuqq}{rgb}{0,0.39215686274509803,0}
\definecolor{zzttqq}{rgb}{0.6,0.2,0}
\definecolor{ttqqqq}{rgb}{0.2,0,0}
\definecolor{qqttqq}{rgb}{0,0.2,0}
\definecolor{qqccqq}{rgb}{0,0.8,0}
\usepackage{tikz-cd}
\usepackage{amsmath}
\usepackage{amsbsy}

\theoremstyle{plain}

\tikzset{
pattern size/.store in=\mcSize, 
pattern size = 5pt,
pattern thickness/.store in=\mcThickness, 
pattern thickness = 0.3pt,
pattern radius/.store in=\mcRadius, 
pattern radius = 1pt}
\makeatletter
\pgfutil@ifundefined{pgf@pattern@name@_b8864pxwu}{
\pgfdeclarepatternformonly[\mcThickness,\mcSize]{_b8864pxwu}
{\pgfqpoint{0pt}{0pt}}
{\pgfpoint{\mcSize+\mcThickness}{\mcSize+\mcThickness}}
{\pgfpoint{\mcSize}{\mcSize}}
{
\pgfsetcolor{\tikz@pattern@color}
\pgfsetlinewidth{\mcThickness}
\pgfpathmoveto{\pgfqpoint{0pt}{0pt}}
\pgfpathlineto{\pgfpoint{\mcSize+\mcThickness}{\mcSize+\mcThickness}}
\pgfusepath{stroke}
}}
\makeatother

\tikzset{
pattern size/.store in=\mcSize, 
pattern size = 5pt,
pattern thickness/.store in=\mcThickness, 
pattern thickness = 0.3pt,
pattern radius/.store in=\mcRadius, 
pattern radius = 1pt}
\makeatletter
\pgfutil@ifundefined{pgf@pattern@name@_f6j18z8qd}{
\pgfdeclarepatternformonly[\mcThickness,\mcSize]{_f6j18z8qd}
{\pgfqpoint{0pt}{0pt}}
{\pgfpoint{\mcSize+\mcThickness}{\mcSize+\mcThickness}}
{\pgfpoint{\mcSize}{\mcSize}}
{
\pgfsetcolor{\tikz@pattern@color}
\pgfsetlinewidth{\mcThickness}
\pgfpathmoveto{\pgfqpoint{0pt}{0pt}}
\pgfpathlineto{\pgfpoint{\mcSize+\mcThickness}{\mcSize+\mcThickness}}
\pgfusepath{stroke}
}}
\makeatother
\tikzset{every picture/.style={line width=0.75pt}} %set default line width to 0.75pt

\begin{document}

\title{Should we necessarily treat masses as localized when analysing tests of quantum gravity?}
\author{Adrian Kent}
\affiliation{Centre for Quantum Information and Foundations, DAMTP, Centre for Mathematical Sciences, University of Cambridge, Wilberforce Road, Cambridge CB3 0WA, UK}
\affiliation{Perimeter Institute for Theoretical
	Physics, 31 Caroline Street North, Waterloo, ON N2L 2Y5, Canada.}
\date{\today}

\begin{abstract}
Recently proposed ``table-top tests of quantum gravity" involve creating, separating
and recombining superpositions of masses at non-relativistic speeds.  
The general expectation is that these generate superpositions of gravitational
fields via the Newtonian potential.   Analyses suggest that negligible 
gravitational radiation is generated if the interference experiments involve
sufficiently small accelerations.   One way of thinking about this is that
matter and the static gravitational field are temporarily entangled and then
disentangled.   Another is that the static gravitational field degrees of freedom
are dependent on the matter and do not belong to a separate Hilbert space, 
and that there is always negligible entanglement between matter and dynamical gravitational degrees of freedom. 

In this last picture, localized masses effectively become infinitely extended objects, inseparable from their Newtonian potentials.    
While this picture appears hard to extend to a fully relativistic theory of non-quantum gravity, it has significant implications for analyses of how or whether BMV and other non-relativistic experiments might test the quantum nature of gravity.   If the masses in a BMV experiment are regarded as occupying overlapping regions (or indeed all of space), explaining how they become entangled
does not require that their gravitational interaction involves quantum information exchange. 
On this view, while the experiments test gravity in a regime where
quantum theory is needed to describe all the relevant matter degrees of freedom, 
they do not necessarily test its quantum nature.   It might be argued that no 
plausible explanation other than quantum gravity could be consistent both with these 
experiments and with special and general relativity.    But any such argument relies on 
further theoretical assumptions, and so is weaker than claiming direct evidence 
for quantum gravitational interactions from the experiments alone.  
\end{abstract}
\maketitle

\section{Introduction}

Proposals by Bose et al. \cite{Bose2017} and Marletto-Vedral \cite{Marletto2017} (BMV) for experiments that could 
entangle two mesoscopic particles gravitationally have attracted a great deal of interest, along with further proposals (e.g. \cite{Howl2021}) for ``table-top 
tests of quantum gravity''.   
Refined versions of BMV experiments, in which entanglement is tested 
by Bell experiments rather than by entanglement witnesses, have been
proposed \cite{Kent2021a} to exclude the possibility that the appearance
of entanglement is explained by local hidden variables.
Table-top tests are very strongly motivated whether or not 
they could definitively confirm quantum gravity, since they would certainly distinguish 
the predictions of non-relativistic quantum mechanics with Newtonian potentials from some
interesting alternatives, including gravitationally-induced collapse models \cite{K66,D84,D87,D89,P96,P98,P14,HPF19} and related ideas \cite{oppenheim2023postquantum} and the possibility that a semi-classical gravity theory holds in the relevant regime \cite{Kent2021b}.   
However, discussion continues over how robustly any of the proposed experiments test the quantum nature of gravity (e.g.\cite{HR18,MV20,H21,HPF19,HVNCRI21,CMT21,martin2023gravity}).   
All the table-top experiments considered to date
probe non-relativistic quantum gravity: the masses move slowly,
and the predictions do not involve $c$.   This is a previously little-explored
regime.  Many physicists' initial reaction is that one cannot get new 
evidence for quantum gravity from predictions derived from the Schr\"odinger
equation with Newtonian potentials.   Yet the prediction that entanglement is 
generated between separated subsystems, when the only significant interaction
is gravitational, does appear at first sight to require a quantumly mediated force. 
It is a basic result of quantum information theory that local operations on and classical communication between separate quantum subsystems cannot generate entanglement.
This is usually framed in a non-relativistic setting, and so apparently applicable
to BMV and related experiments.   
Yet, even prior to considering loopholes, some feel a lingering unease that 
arguments for strong conclusions about quantum gravity are ultimately based on 
no more than standard undergraduate quantum mechanics.   

This paper explores and tries to formalize one version of this unease. 
Its starting point is that arguments that gravity can be shown
to be quantum by analysing only experiments in the non-relativistic
regime need themselves to be consistently non-relativistic.
If we use only non-relativistic dynamical equations to make predictions and
inferences about the underlying physics, 
we should also consider alternative explanations that may not be evidently consistent with relativity.   One reason for this is logical clarity: if in fact we also need to appeal to
relativistic theory or experiment in order to argue that gravity must be quantum,
we should make this part of the argument explicit.   Also, importantly,
such arguments (e.g. \cite{Eppley1977})  sometimes turn out to be incorrect 
\cite{M06,AKR08,K18} and so deserve careful scrutiny.  A lack of evident 
consistency with special or general relativity might possibly be due to a failure of theoretical imagination.  

\section{Non-relativistic charge and mass interferometry}

Consider a particle of charge $q$ in a interferometry experiment with two paths
$\bf{x_L} (t) , \bf{x_R} (t)$ that both
involve small acceleration.  
Suppose that the interference component occupies time $T $, i.e.,
$\bf{x_L} (t) = \bf{x_R} (t)$ for $t\leq 0$ and $t \geq T$, while
$\bf{x_L} (t) \neq \bf{x_R} (t)$ for $0<t<T$.   
In a non-relativistic treatment (effectively taking $c= \infty$) 
its electrostatic field at position ${\bf x_0}$ 
is 
\begin{equation}
{\bf E} ( {\bf x} ) = { q \over { 4 \pi \epsilon_0 }}  {{ { \bf x } - {\bf x_0 }} \over { | {\bf x } - {\bf x_0 } |^3 } } \, .
\end{equation}
During the interval $(0,T)$, the particle's position superposition is associated with
a superposition of electrostatic fields.    
After time $T$ the paths recombine, and there is again a single electrostatic field.
This non-relativistic discussion conceals important subtleties and does not adequately
explain why particle interference is possible in a relativistic world.  
However, it is at least consistent
with the possibility of interference: it implies that, after time $T$, no 
measurement of the electrostatic field can give any path information.
It is in line with the discussion given by Christodoulou and Rovelli \cite{Christodoulou2019}
for the analogous case of non-relativistic BMV mass interferometry. 

A more satisfactory discussion of charge and mass interferometry and relativistic
causality was given by Mari et al. \cite{mari2016experiments}.   
However, we will not discuss their analysis of relativistic causality here,
but rather focus on one point in their illuminating commentary.    As they note, although
the charged particle creates a superposition of electrostatic fields,
there is a sense in which one can treat it as never entangled with 
the quantum electromagnetic field.  The momentum space
Maxwell equation
\begin{equation}
{\bf k}^2 \hat{V} ( {\bf k} ) = \hat{\rho} ({\bf k}) \, 
\end{equation}
gives 
\begin{equation}\label{electrick}
\hat{\bf E} ({\bf k}, t) =  { { - i {\bf k} } \over {{ \bf k }^2 }}  \hat{ \rho} ( {\bf k} ) - {{ \partial} \over {\partial t}}  \hat{ \bf A} ( {\bf k} ) \, . 
\end{equation}
In Coulomb gauge, $ {\bf k} .  \hat{ \bf A} ( {\bf k} ) = 0$ and so 
the longitudinal component of ${\bf E}$ is determined by $\hat{\rho}$
and acts on the same Hilbert space.   The vacuum expectation value of 
the quasistatic electric field is thus determined by that of the charge density
operator.   

As Mari et al. put it, ``the longitudinal component of the electric field is
not a dynamical propagating degree of freedom, since it [$\ldots$] is completely
determined by [external charges], so there is no Hilbert space associated to it.
The Hilbert space of the field contains only the degrees of freedom associated
to the electromagnetic radiation.'' 
In this picture, the matter degrees of freedom can be in a product state with
the field vacuum, despite generating a quasistatic electric field.  
This picture is gauge-dependent, and the Coulomb gauge is not Lorentz invariant. 
But this is not an immediate issue when considering BMV-type experiments in the non-relativistic ($c \rightarrow \infty$) limit and the issue of whether they give evidence for non-relativistic quantum interactions. 

This picture may be counter-intuitive: it seems to suggest
that the electromagnetic field may be in the same vacuum state
whether the particle is localized at ${\bf x}$, ${\bf x'}$
or not present at all.  In this way of thinking, there is nothing
at a general point ${\bf y} \neq {\bf x}, {\bf x'}$.  Yet there {\it is}
an electrostatic field, whose effect is measurable in finite time, 
for example, as Mari et al. discuss, by its action on a charge 
initially trapped in an oscillator near ${\bf y}$.   
One way of reconciling these conflicting intuitions is to think of the
particle as an extended object of which the electrostatic field is an essential 
component: its centre-of-mass wave function and charge density 
may be localized around ${\bf x}$, but the field is not. 

According to this picture, a BMV experiment involving Coulomb
forces can generate entanglement between two particles 
without non-local interactions because the particles -- understood
as extended objects -- always overlap, although their charge densities
do not.   We do not need to appeal to quantum information exchange
to explain the generated entanglement.   
In fact, of course, we have overwhelming evidence 
for QED and in particular that electromagnetic forces {\it are} mediated by 
photons.   We can give a version of this picture based on QED, 
via the unrigorous but still quite compelling 
intuition that we can think of the particle as dressed by an extended 
cloud of virtual electrons, positrons and photons.   
But, importantly, the picture is motivated 
without appealing to an underlying quantum field theory.   

Turning now to non-relativistic gravitational BMV experiments, we need
to keep in mind that their aim is to give evidence for (or against)
the quantum nature of gravity, so we should not require all possible explanations
for the predicted entanglement generation to assume that gravity is
quantized.   We can construct an explanation without
assuming a direct analogue of (\ref{electrick}) and of the underlying theory, QED.
We can simply adopt the analogous picture of a mass $m$ particle as an object
described by non-relativistic quantum mechanics, which has localized centre-of-mass
wave function and mass density, but which has a long-range extension 
associated with its Newtonian gravitational field
\begin{equation}
 - G m ( {{{\bf x } - {\bf x_0 }} \over { |  {\bf x } - {\bf x_0 } |^3  }} )
\end{equation}
This would explain the generation of BMV
entanglement as arising from local interactions between overlapping quantum systems.

A related issue, raised by Hanif et al. \cite{hanif2023testing}, is whether 
decohering a mass interferometer by measuring the path-dependent gravitational
field would constitute evidence for the quantum nature of gravity.
In this picture, it constitutes evidence for the extended nature of a massive
particle, but not that there is an independent quantum gravitational field 
associated with a separate Hilbert space. 
In principle, one could imagine hybrid theories in which matter is extended
in this way, while other gravitational degrees of freedom are classical (or more generally,
not quantum).   

\section{Relativistic considerations}

Considering charges and masses as extended objects, as above, 
is an intrinsically non-relativistic theoretical explanation.
For example, it allows instantaneous action at arbitrary distances. 
This is justified in analysing BMV and other non-relativistic interferometry 
experiments, when comparing standard analyses that do the same. 
Relativistic corrections in these experiments are negligible, 
because all the relevant speeds and distance/time
ratios are small compared to $c$.   In a world in which we had quantum 
mechanics but not relativity, charge and mass interferometry, including 
BMV experiments, would not {\it per se} give compelling evidence that the electromagnetic
and gravitational fields have independent degrees of freedom described 
quantumly, with independent Hilbert spaces, or that they are mediated by
the exchange of quantum particles.   

Of course, classical electromagnetism is a relativistically invariant theory, and we
have compelling evidence for relativistic QED.    We also have compelling evidence
for general relativity.   So our non-relativistic theoretical explanations are 
wrong.    They do, however, pose questions for the interpretation of 
BMV experiments: Could they be extended to relativistically covariant 
explanations consistent with all present experimental data?   
Or to explanations that are not covariant but give
covariant predictions consistent with all present experimental data?   
Or to explanations that are at least consistent
with all present experimental data?   And in each case, can we prove that
the answers are negative, if indeed they are?    

It is worth noting that we can, at least, correct our picture 
to allow for the finite speed of 
propagation of Coulomb or Newtonian potentials. 
For example, in the Newtonian case, consider a slow-moving (i.e. non-relativistic)
particle has trajectory $x(t)$ in the laboratory rest frame.   We take the associated potential at point $y$
and time $t'$ to be 
\begin{equation}\label{causalp}
\Phi (y, t') = { - { G m} \over { | y - x(t_0 ) | }} \, ,
\end{equation}
where $t_0$ is determined by the constraint $t_0 < t'$ and 
\begin{equation}
   { { | y - x(t_0 |} \over { |t' - t_0 |}} = c  \, . 
\end{equation}

In the case of electromagnetism, trying to develop this project towards a 
fully relativistic model, or to 
show that it cannot be so developed, seems
unmotivated.   The totality of evidence for QED is compelling evidence
that the electromagnetic force is quantum.   
BMV experiments to test entanglement generation via Coulomb forces are
technologically interesting but not expected to give new fundamental insights.  
In some sense, QED {\it does} relativistically extend our model, but it does so
in a way that replaces potentials by quantum field excitations.

The case of gravity, however, is presently less clear.    Most theorists 
strongly believe that making our model consistent with quantum theory (for matter)
means replacing it by a quantum theory of gravity.   But the purported point of
BMV-type experiments is to replace this strong belief by conclusive experimental
evidence.    Most theorists (including us) also strongly 
believe that making the classical limit of our model consistent with large-scale
observational data necessitates effectively replacing it by general relativity.
But if so, the case that entanglement generation in BMV experiments implies
the quantum nature of gravity needs to go beyond non-relativistic analyses
and make explicit relativistic assumptions and/or arguments from 
other data. 

\bigskip

\section{Acknowledgements}
I acknowledge financial support from the
UK Quantum Communications Hub grant no. 
EP/T001011/1. 
This work was supported in part by Perimeter Institute for
Theoretical Physics. Research at Perimeter Institute is supported by
the Government of Canada through the Department of Innovation, Science
and Economic Development and by the Province
of Ontario through the Ministry of Research, Innovation and Science.
I am very grateful to Carlo Rovelli for many invaluable comments
and insights.   

\bigskip

\bibliographystyle{unsrt}
\bibliography{library,
postdocbiblioshort}

\begin{thebibliography}{10}

\bibitem{Bose2017}
Sougato Bose, Anupam Mazumdar, Gavin~W. Morley, Hendrik Ulbricht, Marko Toroš, Mauro Paternostro, Andrew~A. Geraci, Peter~F. Barker, M.~S. Kim, and Gerard Milburn.
\newblock Spin entanglement witness for quantum gravity.
\newblock {\em Physical Review Letters}, 119, 2017.

\bibitem{Marletto2017}
Chiara Marletto and Vlatko Vedral.
\newblock Gravitationally induced entanglement between two massive particles is sufficient evidence of quantum effects in gravity.
\newblock {\em Physical Review Letters}, 119, 2017.

\bibitem{Howl2021}
Richard Howl, Vlatko Vedral, Devang Naik, Marios Christodoulou, Carlo Rovelli, and Aditya Iyer.
\newblock Non-gaussianity as a signature of a quantum theory of gravity.
\newblock {\em PRX Quantum}, 2, 2021.

\bibitem{Kent2021a}
Adrian Kent and Damián Pitalúa-García.
\newblock {Testing the nonclassicality of spacetime: What can we learn from Bell–Bose et al. -Marletto-Vedral experiments?}
\newblock {\em Physical Review D}, 104, 2021.

\bibitem{K66}
F.~Karolyhazy.
\newblock Gravitation and quantum mechanics of macroscopic objects.
\newblock {\em Nuovo Cimento A (1965-1970)}, 42:390--402, 1966.

\bibitem{D84}
L~Di{\'o}si.
\newblock Gravitation and quantum-mechanical localization of macro{-}objects.
\newblock {\em Phys. Lett. A}, 105(4-5):199--202, 1984.

\bibitem{D87}
L.~Di\'{o}si.
\newblock A universal master equation for the gravitational violation of quantum mechanics.
\newblock {\em Phys. Lett. A}, 120:377--381, 1987.

\bibitem{D89}
L.~Di\'osi.
\newblock Models for universal reduction of macroscopic quantum fluctuations.
\newblock {\em Phys. Rev. A}, 40:1165--1174, 1989.

\bibitem{P96}
R.~Penrose.
\newblock On gravity's role in quantum state reduction.
\newblock {\em Gen. Relat. Gravit.}, 28:581--600, 1996.

\bibitem{P98}
Roger Penrose.
\newblock Quantum computation, entanglement and state reduction.
\newblock {\em Phil. Trans. R. Soc. A.}, 356(1743):1927--1939, 1998.

\bibitem{P14}
Roger Penrose.
\newblock On the gravitization of quantum mechanics 1: Quantum state reduction.
\newblock {\em Found. Phys.}, 44(5):557--575, 2014.

\bibitem{HPF19}
Richard Howl, Roger Penrose, and Ivette Fuentes.
\newblock Exploring the unification of quantum theory and general relativity with a {B}ose{\textendash}{E}instein condensate.
\newblock {\em New J. Phys.}, 21(4):043047, apr 2019.

\bibitem{oppenheim2023postquantum}
Jonathan Oppenheim.
\newblock A postquantum theory of classical gravity?
\newblock {\em Physical Review X}, 13(4):041040, 2023.

\bibitem{Kent2021b}
Adrian Kent.
\newblock Quantum state readout, collapses, probes, and signals.
\newblock {\em Physical Review D}, 103, 2021.

\bibitem{HR18}
Michael J~W Hall and Marcel Reginatto.
\newblock On two recent proposals for witnessing nonclassical gravity.
\newblock {\em J. Phys. A: Math. Theor.}, 51(8):085303, jan 2018.

\bibitem{MV20}
Chiara Marletto and Vlatko Vedral.
\newblock Witnessing nonclassicality beyond quantum theory.
\newblock {\em Phys. Rev. D}, 102:086012, Oct 2020.

\bibitem{H21}
Simon~A Haine.
\newblock Searching for signatures of quantum gravity in quantum gases.
\newblock {\em New J. Phys.}, 23(3):033020, mar 2021.

\bibitem{HVNCRI21}
Richard Howl, Vlatko Vedral, Devang Naik, Marios Christodoulou, Carlo Rovelli, and Aditya Iyer.
\newblock Non{-}gaussianity as a signature of a quantum theory of gravity.
\newblock {\em PRX Quantum}, 2:010325, Feb 2021.

\bibitem{CMT21}
Daniel Carney, Holger M\"uller, and Jacob~M. Taylor.
\newblock Using an atom interferometer to infer gravitational entanglement generation.
\newblock {\em PRX Quantum}, 2:030330, Aug 2021.

\bibitem{martin2023gravity}
Eduardo Mart{\'\i}n-Mart{\'\i}nez and T~Rick Perche.
\newblock What gravity mediated entanglement can really tell us about quantum gravity.
\newblock {\em Physical Review D}, 108(10):L101702, 2023.

\bibitem{Eppley1977}
Kenneth Eppley and Eric Hannah.
\newblock The necessity of quantizing the gravitational field.
\newblock {\em Foundations of Physics}, 7, 1977.

\bibitem{M06}
James Mattingly.
\newblock Why {E}ppley and {H}annah's thought experiment fails.
\newblock {\em Phys. Rev. D}, 73:064025, Mar 2006.

\bibitem{AKR08}
Mark Albers, Claus Kiefer, and Marcel Reginatto.
\newblock Measurement analysis and quantum gravity.
\newblock {\em Phys. Rev. D}, 78:064051, Sep 2008.

\bibitem{K18}
Adrian Kent.
\newblock Simple refutation of the {E}ppley{\textendash}{H}annah argument.
\newblock {\em Class. Quantum Grav.}, 35(24):245008, nov 2018.

\bibitem{Christodoulou2019}
Marios Christodoulou and Carlo Rovelli.
\newblock On the possibility of laboratory evidence for quantum superposition of geometries.
\newblock {\em Physics Letters, Section B: Nuclear, Elementary Particle and High-Energy Physics}, 792, 2019.

\bibitem{mari2016experiments}
Andrea Mari, Giacomo De~Palma, and Vittorio Giovannetti.
\newblock Experiments testing macroscopic quantum superpositions must be slow.
\newblock {\em Scientific reports}, 6(1):22777, 2016.

\bibitem{hanif2023testing}
Farhan Hanif, Debarshi Das, Jonathan Halliwell, Dipankar Home, Anupam Mazumdar, Hendrik Ulbricht, and Sougato Bose.
\newblock Testing whether gravity acts as a quantum entity when measured.
\newblock {\em arXiv preprint arXiv:2307.08133}, 2023.

\end{thebibliography}

\end{document}